\pdfoutput=1 
\documentclass[english,nohyper,a4paper]{tufte-handout}
\usepackage{amsmath,amssymb,amsthm}

\usepackage[latin9]{inputenc}
\usepackage[T1]{fontenc}
\usepackage{babel}

\usepackage{color}
\usepackage{wasysym}   
\usepackage{marvosym}  
\usepackage{graphicx}
\usepackage{microtype}
\def\adx#1:#2\par{\par\halign{\hskip #1##\hfill\cr #2}\par}

\def\teff{T_{\rm eff}}
\def\mast{M_\ast}

\def\msol{M_\odot}

\def\llsol{L/L_\odot}

\def\sigr{\sigma_{\rm R}}
\def\sigi{\sigma_{\rm I}}

\def\diff{{\mathrm d}}
\def\Diff{{\mathrm D}}
\def\unity{ \hbox{1\kern-.23em l} }
\def\zero{ \hbox{0\kern-.23em |} }
\def\field{ \hbox{I\kern-.23em K} }

\def\braket #1.#2.{\langle #1 \vert #2 \rangle}

\newcommand{\lyxaddress}[1]{ \vspace{1.4em} \par {\raggedright #1 \vspace{1.4em}
\noindent\par} }

\usepackage{amsmath}
\title{The theorem that was none - II. The profound 70s} 
\author{Alfred Gautschy}
\date{}

\begin{document} \maketitle  
\lyxaddress{CBmA, 4410 Liestal, Switzerland}


\marginnote[+0.8cm]{From \citet{Lauterborn1976a}} \begin{quotation}
$\left[\dots\right]$ \emph{One of the basic but essentially unsolved
                     problems in stellar structure theory is to
                     understand the transition from local to global
                     properties of stars.} 
$\left[\dots\right]$ \end{quotation}

\section*{Introduction}
The Vogt-Russell (VR) theorem states, crudely speaking, that a
star's mass and its chemical composition define uniquely its structure
and hence its position in any suitably chosen characteristic diagram,
such as the HR~--, mass~-~radius~--, central-density~-~temperature~--~diagram. 
The simple claim of the theorem
seems, however, to be contradicted, at least by model stars. In any
case, discussions of the VR theorem force astrophysicists to
think thoroughly about the various approximations going into the
modeling of stars, about limitations of numerical modeling, and the
mathematical properties of the involved differential equations.   

A first installment \citep{Gautschy2015} focused on the early years 
of the theorem and its reception to the end of the 1960s. The
exposition stopped at about the time when the computational stellar
evolution industry took off. The following second part of the
historical discourse on the VR theorem focuses mainly on achievements
during 1970s. Those years constituted the first phase 
of extensive stellar-evolutionary computations at various levels 
of abstraction. The advanced evolutionary stages of star
models reached thereby revealed evidence of violations 
of the VR theorem. Parallel to all the computational work, 
more abstract mathematical methods were imported to study the 
solution properties of the stellar structure equations. 
During the early 1970s, a small number of researchers were attracted
to the problem of the validity of the VR theorem. After the mid 1970s, 
however, that kind of formal stellar-physics research died out again. 
The quite formal and detached findings regarding the behavior 
of solutions of the stellar-evolution equations remained confined to
a small group of researchers; but most importantly, the insights did 
not diffuse into the general literature and did not influence 
textbook opinions~--~with the self-explanatory exception of 
\citet{kw}. After the late 1970s, stellar-evolution research returned 
to and remained essentially a numerical-computation enterprise 
focusing on data-driven modeling.

\begin{marginfigure}[-9.0cm] 
	\begin{center}
	\includegraphics[width=0.97\textwidth]{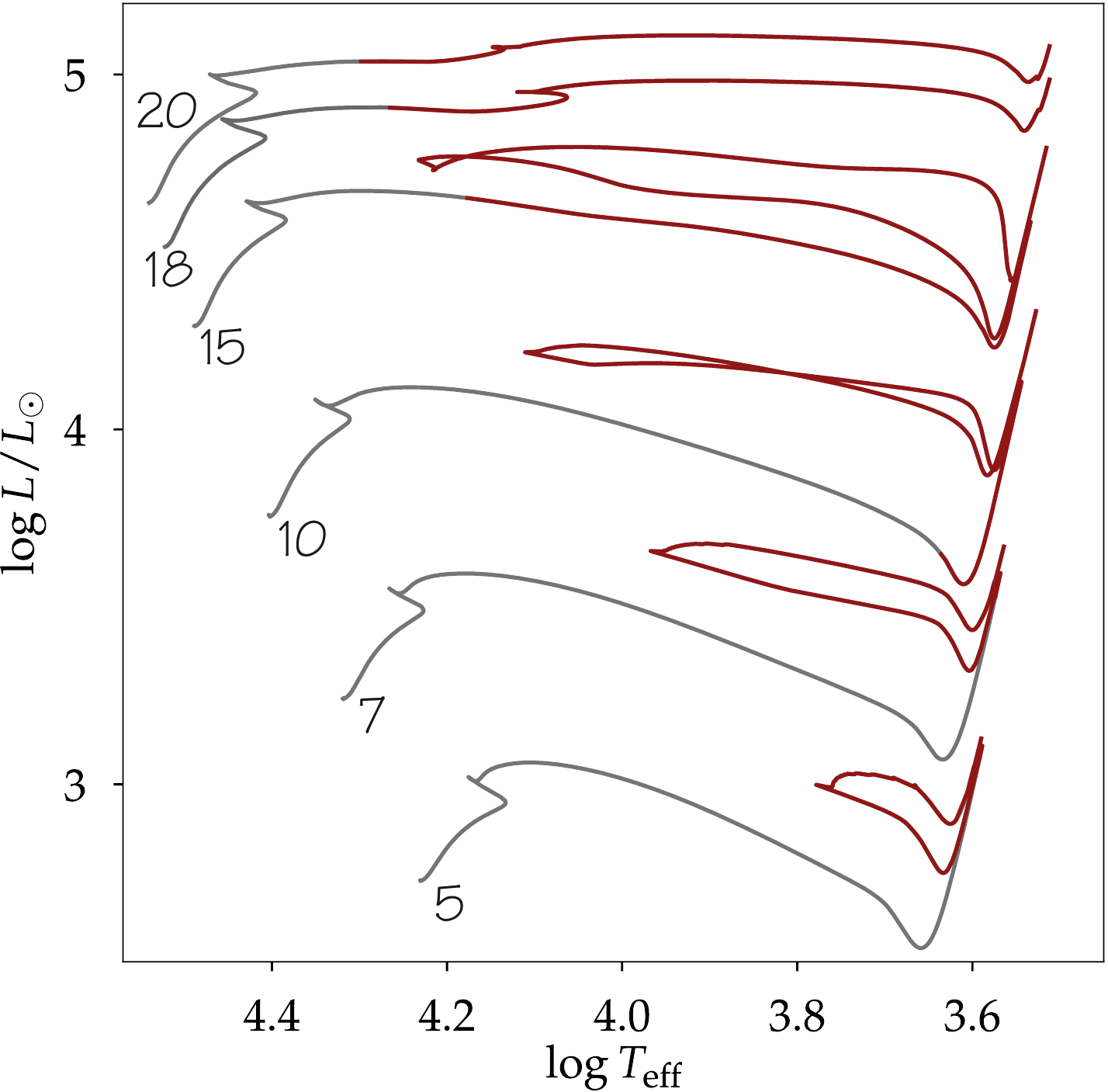} 
	\end{center}
     \caption{ Evolutionary tracks of intermediate-mass and massive
               stars through hydrogen (grey lines) and helium burning
               ($L_{3\alpha}>L_\odot$; red lines). 
               Masses in solar units are noted close to the
               ZAMS loci of the respective tracks. 
             } \label{fig:AllBlueLoops}
\end{marginfigure}
\section*{Counterexamples and conjectures} 
Analyzing the results from his own stellar-evolution modeling and
comparing them with data from the published literature,
\citet{Paczynski1970} was confronted with annoying divergences in size
and partly even the existence of blue loops traced out on the HR
diagram by intermediate and massive stars (3, 5, 7, 10, and 15 $\msol$) 
during their core helium-burning
phase.\sidenote[][-1cm]{Illustrative evolutionary paths of 
		                intermediate-mass and massive stars, as computed with
		                the MESA code, are shown
		                in Fig.~\ref{fig:AllBlueLoops}.
	                    Blue loops during core helium burning are
	                    experienced by stars up to about $15 \msol$.}
After he failed to disclose any sensible correlations of model
parameters with the bluest points on the HR plane reached during the
blue loops of $5 \msol$ stars, Paczy{\'n}ski conjectured that multiple
solutions and associated thermal instabilities might be the source of
the problem, at least for the more massive  
stars.\sidenote[][0.0cm]{Ironically enough, the $5 \msol$ case usually 
                 turns out to be thermally stable during its blue loop; 
                 this means that the \emph{existence }of a blue loop does 
                 not depend on a secular instability of the star. This had 
                 become evident already early on in
				 \citet{LauterbornRefsWeigert71}.}
The effect of the thermal instability was thought to be sensitive to
the numerical treatment of the subphotospheric layers, therefore,
Paczy{\'n}ski concluded that the different numerical treatments of the
various authors as well as his own ones affected the eventual expression
of the blue loops. Even though Paczy{\'n}ski did not mention the VR
theorem explicitly, his suspicion of multiple solutions and the 
particular phrasing of the text must  
have meant to him that the VR theorem was apparently violated.
\citet{Kozlowski1971}, a young researcher advised by Paczy{\'n}ski, looked
closer into the blue-loop problem by constructing linear series of
full-equilibrium (FE) models of $10 \msol$ stars. In his 
\emph{fitted shooting models}, he found that multiple solutions to the
full-equilibrium stellar structure equations constructed for helium-core
masses in the range  of
$2.43 < \msol < 2.53$.\sidenote{The
  \emph{abundance profile} was identified as being important; cool
  star models could not be recovered with step profiles alone. Only
  after introducing more realistic ramp structures, blue and red
  branches of equilibrium solutions emerged. The detailed structure of
  the abundance profile around the H-burning shell turned out to be
  crucial in the whole discussion of the cause of blue-looping stars.}
The multiple solutions of equal helium-core mass models mimicked
indeed the blue loops on the HR plane of the full stellar-evolution models.

In a series of six papers published between 1972 and 1977, 
Paczy{\'n}ski and his various co-authors further explored 
the solution behavior of linear series of FE star models. The second paper,
\citet{Paczynski1972}, was devoted to homogeneous pure carbon stars,
for which the total stellar mass was adopted as control parameter.  At the very
least, a stable main-sequence~--~like (non degenerate) solution family
and a stable degenerate sequence, reminiscent of carbon white dwarfs
were encountered.  When neutrino losses were added to the
computations, the authors reported hints of additional solution
sequences, possibly thermally unstable ones, which at the time turned
out to be very tricky to track numerically. 
These sequences were characterized by the number of 
regions with negative total luminosity therein; and this number remained
invariant under change of the control parameter. Therefore,
the authors conjectured that large number of
sequences might lurk in solution space and hence, a correspondingly
severe violation of the VR theorem might prevail. The solution
branches with these multiple negative luminosity regions were, to the
best of my knowledge, never followed up; so existence and relevance of
such models remains obscure. Paper three of the series,
\citep{Kozlowski1973}, gained some fame for its
linear series of $10 \msol$ star models with inconspicuous
hydrogen/helium~--~envelopes that mimicked their core-helium burning
phase. FE models with helium core masses in the range $0.37 - 0.375 \msol$
sported \emph{nine }simultaneous equilibrium solutions.

\smallskip

Around 1970, Alfred Weigert started a small research group at Hamburg
Observatory. Using, what was colloquially referred to as the
Kippenhahn stellar-evolution code and to which Weigert was a founding
contributor, the Hamburg group tackled various aspects of single- and
binary-star evolution. Dietmar Lauterborn, then a young researcher in
Weigert's group, studied~--~independently of Paczy{\'n}ski's quests~--~the
\emph{physical cause} of the blue loops of helium-burning,
intermediate-mass (for $3 \msol$ and mostly $5 \msol$)
stars \citep{LauterbornRefsWeigert71} to grasp the bewildering variety
of loci of such stars on the HR
plane.\sidenote[][]{As early as 1964, blue loops of
  intermediate-mass stars had been sighted in evolutionary
  computations performed with the "Kippenhahn code"
  \citep{Hofmeister1964} and also in results obtained with 
  the "Iben code" \citep{Iben1964} }
The stellar models discussed in the paper relied on the generalized
main-sequence approach of \citet{Giannone1968} where 
FE models were computed with the Kippenhahn code 
adopting prescribed composition profiles. The parameterized mass 
of the helium core mimicked thereby the temporal evolution of 
the model sequence. 

The first paper of \citet{LauterbornRefsWeigert71} did not yet
mention multiple solutions and secularly unstable model branches
were not involved. Nonetheless, the paper left its mark and
influenced the blue-loop literature by the analytical prescription
of the H/He profile as a truncated ramp around the H-burning
shell, this shape favors blue excursions of the respective
model stars. The authors took advantage of the fact that the position 
on the HR plane of a centrally He-burning model star depended essentially
on the \emph{macroscopic }properties of the helium core. 
The detailed structure of the core is not important, only core mass, 
core radius, and the luminosity passing through the core's surface were 
identified to matter. Specifying these core quantities as inner boundary
conditions of envelope-only computations granted a great deal of
freedom to construct envelope sequences for which one 
or several of the core properties (entering as inner boundary conditions) 
could be varied at discretion. In particular the physical properties 
of the \emph{bottoms of envelopes }(BoE) computed in this artificial 
way provided crucial arguments on the issue of multiple solutions 
to the stellar structure equations. With this approach of splitting 
up a model star into a core and an attached envelope, the Henyey relaxation 
approach regained characteristics otherwise typical of two-sided 
shooting methods for boundary-value problems.
    
\begin{marginfigure} 
	\begin{center}
	\includegraphics[width=0.97\textwidth]{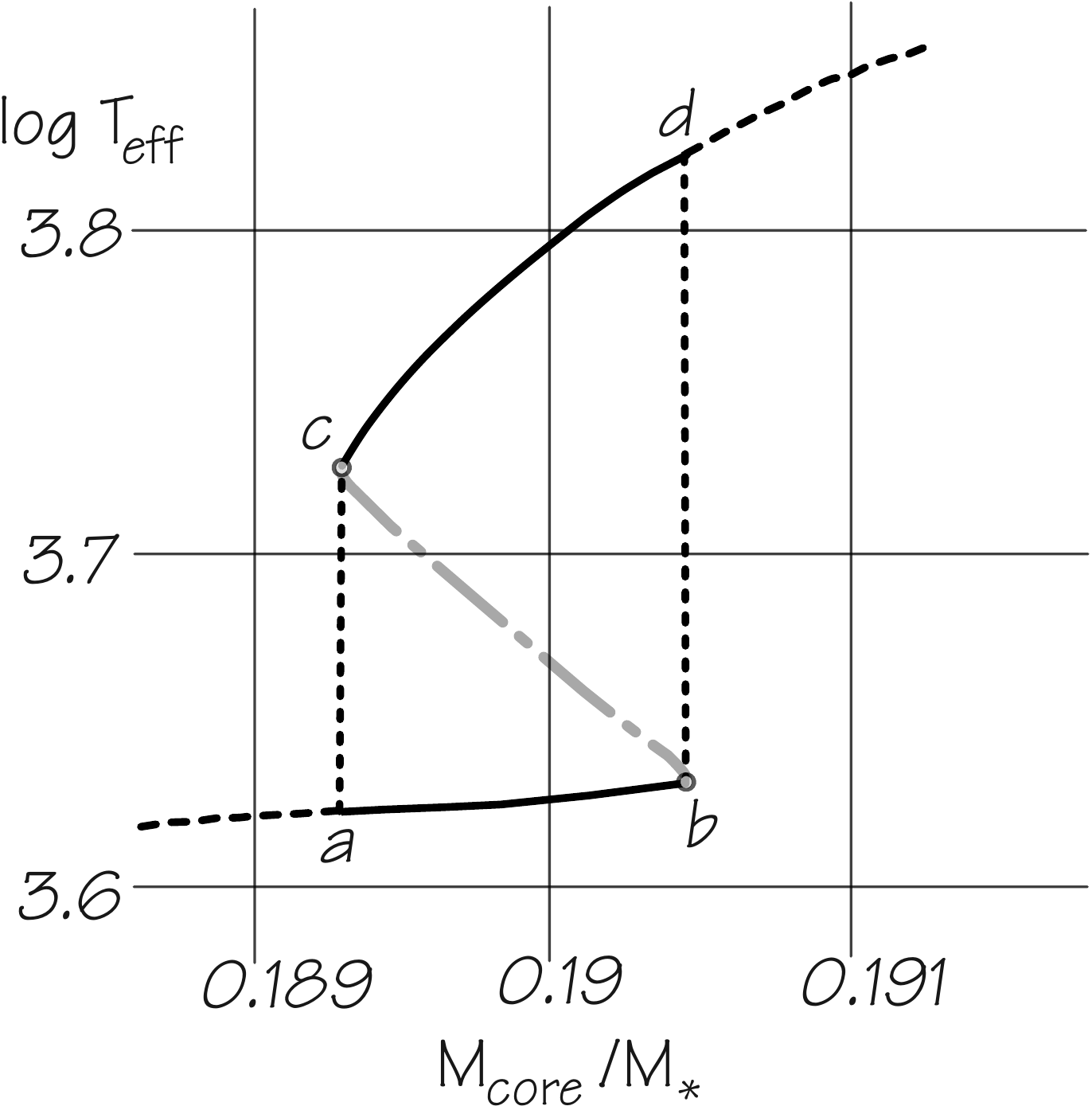} 
	\end{center}
     \caption{Core-mass~--~effective temperature relation for a 
              blue-looping $9 \msol$ model sequence after 
              \citet{LauterbornRefsRoth71}. 
             } \label{fig:Fig7b}
\end{marginfigure}

In the second paper \citep{LauterbornRefsRoth71}, which addressed 
more massive stars at $7$ and $9 \msol$, thermal
instability and multiplicity of solutions entered the stage.
Figure~\ref{fig:Fig7b} sketches the effective temperatures as a
function of helium core mass of the $9 \msol$ FE
sequence of the complete star models from \citet{LauterbornRefsRoth71}. 
Two solution branches were encountered: As the helium core mass grows,
a cool branch evolves at low effective temperatures, going through
point \emph{a }to point \emph{b }where it terminates.  Starting with the helium
core mass of point \emph{b}, a second, higher-temperature sequence 
exists that passes through point \emph{d}. As the helium core
grows,\sidenote[][-1.4cm]{This means that hydrogen shell-burning drives the
                star's evolution. The effect of the helium core-burning
                luminosity and the associated composition change 
                can be neglected in this case.}
the star `evolves'~--~in full equilibrium~--~to the upper right of the
diagram. Again, starting at point \emph{d}, the high-$\teff$ solution branch
of FE star models can be  followed to point \emph{c }upon lowering the
helium-core mass where the equilibrium series terminates again. 
If the helium core mass is further reduced,
\emph{no }FE model star can be found in the direct neighborhood of point \emph{c}.
In the tiny range
$0.1893 < M_{\mathrm{He\,core}}/\mast < 0.1905$ the two sequences
overlap, i.e. two equilibrium solutions exist per prescribed
helium core mass. Actually, points \emph{b }and \emph{c }can be connected by an
FE solution branch (dash-dotted line hinted at
schematically in Fig.~\ref{fig:Fig7b}), which is secularly 
unstable.
For an actual star, the double- or actually triple-solution episode
never causes a problem. Evolution and the appropriate physical
processes resolve the di- or better trilemma: A star would approach 
point \emph{a }as the helium core grows
and it would~--~essentially in full equilibrium continue its evolution
until the equilibrium sequence terminates at point \emph{b}. Because the
helium core of an actual star can only grow at this point, 
the star adapts to the circumstances by going thermally unstable 
and evolving~--~quickly compared with FE 
evolution~--~in the vicinity of line \emph{b}-\emph{d} to find new full equilibrium
states at around epoch \emph{d}. Thereafter, the star continues its evolution 
at higher effective temperature, again close to a full-equilibrium locus, 
hinted at by the dashed line to the upper right of the
figure.\sidenote{Intermediate-mass and massive stars live through a 
comparable situation when they enter their Sch\"onberg-Chandrasekhar 
instability phase.}

After \citet{LauterbornRefsWeigert71}, it was evident that blue loops
of core helium-burning stars could develop also in full equilibrium
models, at least for not too massive stars. With growing total stellar
mass, FE tends to get lost and secular instability
develops as the helium core grows, and the magnitude (measured in the
effective-temperature range they swept on the HR plane) of the loops
grows. Hence, the existence of blue loops does not depend on
secular instability and therefore blue loops are not causally related to
uniqueness issues of stellar models. 

\smallskip March 1972 must have been a busy period at Hamburg
Observatory: Within 18 days, three papers on the same topic were
submitted for publication by different members of Weigert's research
group (although Lauterborn apparently was on leave of absence at
JILA in Boulder, most likely with John P.~Cox). 
The editorial office of A\&A received the
\citet{Lauterborn1972} manuscript on March 2nd, that of
\citet{Kaehler1972} on March 3rd, and finally the \citet{Roth1972} one
on March~20th. 

\citet{Lauterborn1972} set out to understand the BoE loci and those 
traced out by the surfaces of core solutions on the radius~--~pressure 
plane.\sidenote[][-0.5cm]{The core was defined as the volume of the star 
  where $X = 0$; the rest of the mantle with $X>0$ was attributed to 
  the envelope; i.e. the H-burning shell was considered as the bottom
  part of the stars' envelopes.} 
The computations revealed that core and envelopes loci produced up to three
intersections and therefore gave rise to triple degeneracies for some
cases of equilibrium-structure solutions of blue-looping supergiants. 
Section~3 of the Lauterborn paper eventually addressed the
violation of the VR theorem. In a terse paragraph, he kept 
the ball low when he emphasized that the correct interpretation of the
classical VR theorem be a local, not a global one~--~i.e. only one 
solution may exist in a suitably chosen \emph{neighborhood} of a 
given solution. Lauterborn did not dive deeper into mathematical
technicalities but focused on the particular class of stars at hand and he
studied how the solutions depended on the numerous physical
parameters that usually enter the modeling of these stars. A bump
(also referred to as the \emph{hook} in pertinent papers) in the
BoE loci\sidenote{In case of the Sch\"onberg-Chandrasekhar instability, 
it is the looping loci of the isothermal helium \emph{core }solutions 
in characteristic fitting diagrams that give rise to multiple equilibrium 
solutions \citep[see e.g.][]{Gabriel1967, Roth1972}.}
on the pressure~--~radius plane emerged as being the
reason of potential multiplicity of the solutions to the FE
stellar structure problem.  The physical origin of the
non-monotonicity was, however, not yet elaborated. 

The \citet{Roth1972} paper expanded on the multiple solutions of
FE models with pure helium cores and hydrogen-rich envelopes. 
The authors reviewed earlier papers that dealt with generalized 
main-sequences and emphasized that the double solutions encountered 
in that context were caused by the different material properties of 
helium cores that could successfully match envelope solutions: 
One solution branch was made up by non-degenerate He-burning cores, 
the other consisted of degenerate, isothermal cores. \citet{Roth1972}
contrasted the \citet{LauterbornRefsWeigert71} multiple solutions found 
for massive stars emphasizing that the multiple solutions were caused 
by the particular material properties of the envelopes rather than
the cores. Roth et al. managed to 
compute much broader radius and pressure ranges on the core~--~envelope
fitting plane. This allowed them to recover not only the "hook"
localized by \citet{LauterbornRefsWeigert71}; they also tracked down 
three additional solution possibilities of models with isothermal cores. 
Therefore, the triple solutions of Lauterborn et al. were actually 
sextuple solutions. 

For solutions to FE model stars with isothermal helium 
cores, Roth \& Weigert identified the roots that are associated with the 
Sch\"onberg-Chandrasekhar instability. Most interestingly, the authors
outlined concisely the topological behavior of the roots at the onset of the 
secular instability. Using the concept of the 
Henyey-determinant~--~which was elaborated on in  
\citet{Paczynski1972a} and \citet{Kaehler1972}~--~the authors pointed out 
that zeros of the determinant indicate the loss of FE
at the onset of the  triple solutions of the $9 \msol$ star studied previously
by Lauterborn and collaborators.
 
\citet{Murai1974} looked into the multiple-solution problem resorting
to the methods developed earlier on by the Japanese stellar-astrophysics 
school around Hayashi, H{\=o}shi, and Sugimoto. By performing the fitting of
interior and exterior solutions in the plane of the homology
invariants $U$ and $V$, the spread of the fitting quantities on the
fitting plane could be significantly
reduced.\sidenote{The reason is that homologous solutions of 
the stellar structure equations are invariant on the $U$,$V$ plane, 
i.e. they trace the same locus thereon.} 
Murai found the reason of the hook to be tied to the particular opacity behavior
in the deep interior of the envelope models. It is the functional form of 
the Kramers law that induces a spiraling motion on the $UV$-~plane 
of envelope solutions. Effects of radiation pressure and 
of course of convection suppress this spiraling tendency. Consequently, 
stars that are sufficiently hot for their matter to be dominated 
by electron scattering or cool enough to have extended convective envelopes 
lose their ability to allow for multiple solutions at a 
given total mass and a given chemical composition. 

Additionally, Murai pointed out the coincidence of the occurrence of the hook in 
the fitting behavior of envelope models and the development of the Hertzsprung-gap.
Even though earlier authors attributed the 
Sch\"onberg-Chandrasekhar instability to the isothermal spiral of the inert 
helium \emph{core}. Figure~\ref{fig:m050z02} of this paper 
illustrates that for the example of a $5 \msol$ model star with 
pure electron-scattering opacity only (i.e. models with suppressed "hook") 
can be evolved through the Hertzsprung gap in FE, which is 
to say that they do not experience the SC instability anymore. 
  
At a meeting of the \emph{Astronomische Gesellschaft }in 1975, 
\citet{Lauterborn1976a} reviewed
the physical origin of the earlier found triple solutions to massive stars 
in their central helium-burning phase; he made extensive 
use of Murai's $UV$-~plane viewpoint and illustrated the robustness of 
Murai's identification of the Kramers opacity effect as the origin of
the multiple of solutions in the $7 \msol$-star case.

\section*{Local uniqueness}
K\"ahler, another young member of Weigert's research group in Hamburg, 
chose to attack the problem of the VR theorem
more formally without having specific evolutionary scenarios in mind.
In the first paper of \citet{Kaehler1972}, he focused on the behavior 
of FE models. The starting point was the set of the canonical stellar-evolution
equations\sidenote[][]{which can be pruned directly from
	                   equations $2, 6, 8,$ and $10$ in \citet{Gautschy2015}.
	                   These equations are referred to as I.2, I.6, I.8, and I.10
	                   henceforth.}  
with thermodynamic basis $\{P,T\}$; for a particular stellar model, 
the set of parameters, $\wp = \wp\left(\mast ,\vec{\chi} \right)$, 
was assumed to be prescribed. 

\begin{figure} 
	\begin{center}
	  \includegraphics[width=0.95\textwidth]{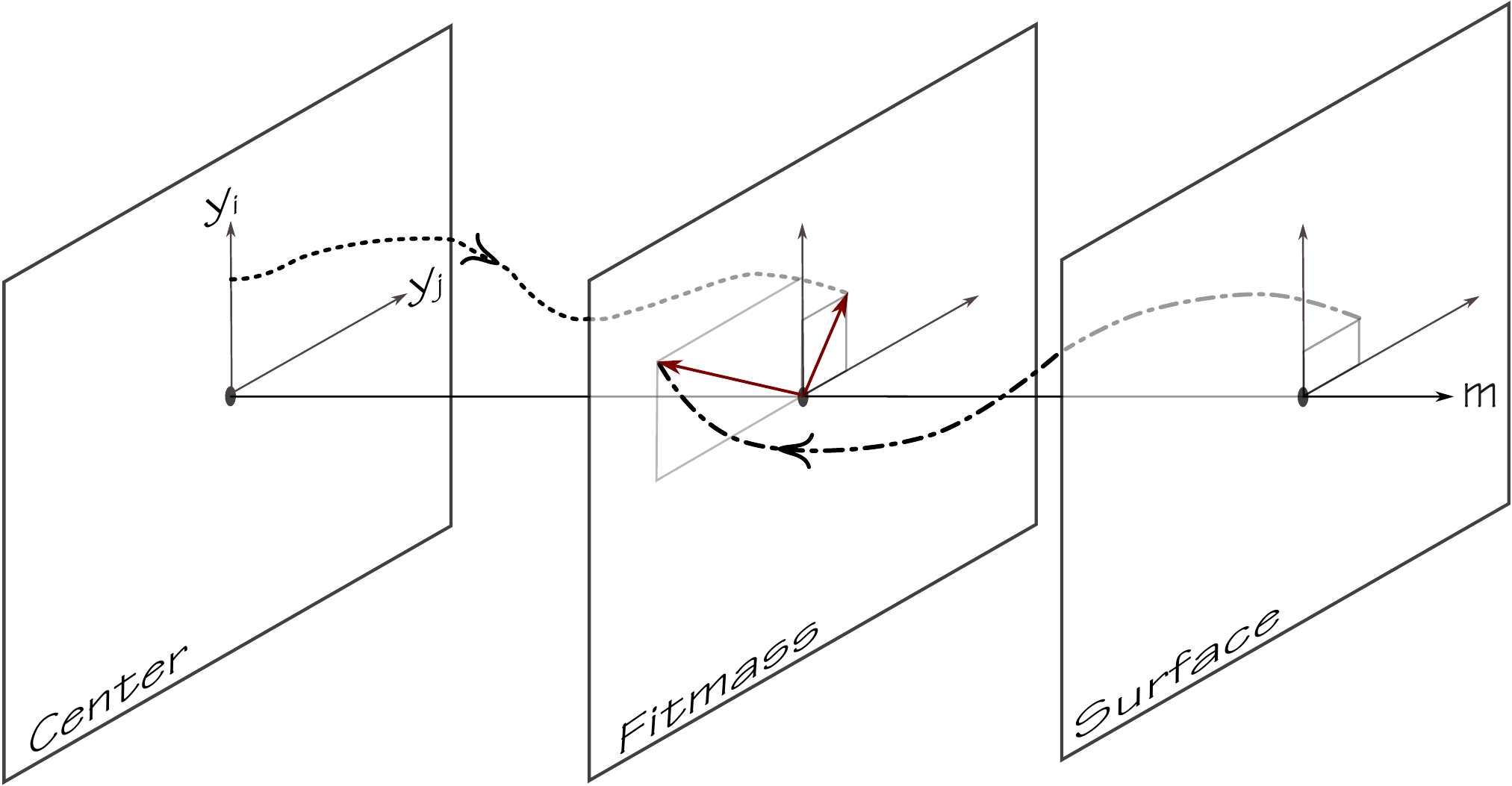}
	\end{center} 
	\caption{Illustration of the IVP strategy to solve 
	         a hypothetical two-sided BVP of two variables.
	         For the stellar structure problem, this procedure
	         operates in a four-dimensional space. The solutions
	         $y_{\mathrm{I}}$ live between the
	         center- and the fitmass-plane; whereas $y_{\mathrm{O}}$
	         unfold between surface- and fitmass-plane.
	        } \label{fig:TwoSided}
\end{figure}

The solution of the FE structure problem was assumed to be obtained with
some two-sided shooting method. In this case, 
the \emph{in-out }integration, starting at the stellar 
center to the to some fitting mass, $m_{\text{F}}$, constitutes 
an initial-value problem, which starts at a regular singularity. The annoyance
of this singularity is usually cushioned by starting the computation 
at some small distance off the center with the solution expanded 
into a power series. Other than that the right-hand sides of the 
differential equations are well behaved so that they can be considered 
locally Lipschitz. Therefore, the existence of a solution (of the IVP) is
guaranteed. The same reasoning applies to the 
\emph{out-in }integration, which starts at a suitable, 
\emph{physically }motivated specification of a stellar surface; 
again, the integration extends to $m_{\text{F}}$.
Figure~\ref{fig:TwoSided} illustrates the two-sided integration strategy. 
Two trial solutions obtained from the in-out and the out-in integrations
meet at $m_{\text{F}}$. Only if they exactly match, and if  they have 
additionally the same derivatives at $m_{\text{F}}$ do they constitute 
a physically acceptable solution. 

The distributed boundary conditions allow the two equations to be
integrated from one side of the problem each; trial values are assumed for
the remaining two boundary conditions: $P_{\text{c}}$ and $T_{\text{c}}$ in the
center and say $L$ and $R$ at the surface. Usually, the magnitudes of the
dependent variables to not match at the fitmass $m_{\text{F}}$, 
cf. the red state vectors at the fitmass plane in 
Fig.~\ref{fig:TwoSided}. Iteratively improving the guesses of the yet 
unknown boundary values will eventually lead to a solution of the
stellar structure problem via a series of IVP integrations.

\begin{figure} 
	\begin{center}
	  \includegraphics[width=0.95\textwidth]{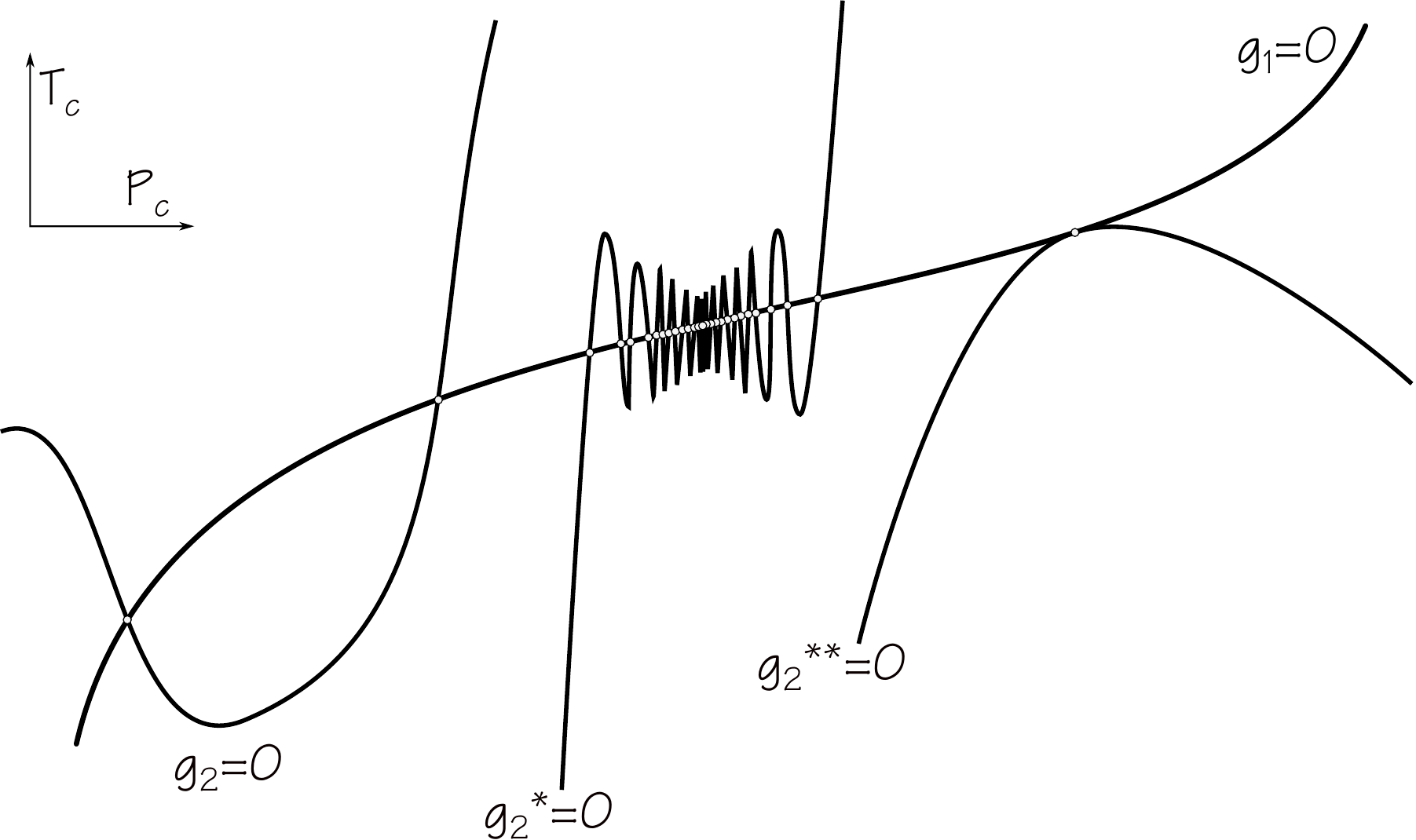}
	\end{center} 
	\caption{Possible nontrivial intersection topologies of the two 
	         fitting conditions on the $P_{\mathrm{c}} - T_{\mathrm{c}}$ 
	         parameter plane. For simplicity, one function
	         $g_1=0$ and different options of the behaviors of $g_2=0$ 
	         are adopted.
	        } \label{fig:IntersectionOptions}
\end{figure}

\citet{Kaehler1972} chose to split up the problem.
He specified surface values for $L$ and $R$ and  
integrated the mass and energy equations (eqs.~I.2 and I.8) down to $m_{\text{F}}$. 
Analogously, these equations can be integrated in-out from the center, 
whereas they depend implicitely on $P_{\text{c}}$ and $T_{\text{c}}$ via 
the quantities $\rho$ and $\varepsilon$. The parameters are iteratively
updated until the solutions match at $m_{\text{F}}$:
\begin{align}
	\label{eq:r_fit}
	r_{\text{I}}(m_{\text{F}};P_{\text{c}},T_{\text{c}}) 
	        & = r_{\text{O}}(m_{\text{F}}; R, L)\, ,             \\
	\label{eq:L_fit}
	L_{\text{I}}(m_{\text{F}};P_{\text{c}},T_{\text{c}}) 
	        & = L_{\text{O}}(m_{\text{F}}; R, L)\, .
\end{align}
Equations (\ref{eq:r_fit}) and (\ref{eq:L_fit}) suggest that the outer boundary
conditions $R$ and $L$ can be related to $P_{\text{c}}$ and $T_{\text{c}}$.
If these relations are monotonous in the physically relevant parameter space then
relations $R(P_{\text{c}},T_{\text{c}}),L(P_{\text{c}},T_{\text{c}})$ can 
be derived. Eventually, the remaining fitting conditions at $m_{\text{F}}$ 
boil down to two  relations that depend on $P_{\text{c}}$ and $T_{\text{c}}$ 
only:
\begin{align}
	\label{eq:g_1}
	g_1 & =   P_{\text{I}}(m_{\text{F}};P_{\text{c}},T_{\text{c}}) 
	        - P_{\text{O}}(m_{\text{F}};R(P_{\text{c}},T_{\text{c}}),L(P_{\text{c}},T_{\text{c}})) \\
    \label{eq:g_2}
	g_2 & = T_{\text{I}}(m_{\text{F}};P_{\text{c}},T_{\text{c}}) - T_{\text{O}}(m_{\text{F}};R(P_{\text{c}},T_{\text{c}}),L(P_{\text{c}},T_{\text{c}}))
\end{align}
The two relations, $g_1$ and $g_2$, illustrate the nature of the 
boundary-value problem. An intersection of $g_1 = 0$ 
\emph{and }$g_2 = 0$ on the $P_{\text{c}} - T_{\text{c}}$ plane means 
that the stellar-structure problem admits of a solution. The behavior of the
two nonlinear equations $g_1=0$ and $g_2=0$ can be complicated with, in principle,
wild twists and turns (e.g. Fig.~\ref{fig:IntersectionOptions}). It is possible 
that $g_1=0$ and $g_2=0$ do not intersect at all; they may intersect once or 
multiple times for a fixed choice of $\wp$; 
this latter case will then be a realization of the infamous 
multiple-solution cases that have been encountered numerically. 
The multiple-solution cases can be manifold,
Fig.~\ref{fig:IntersectionOptions} suggest two solutions with a finite distance
on the $P_{\text{c}} - T_{\text{c}}$ plane ($g_1=0$ and $g_2=0$). If $g_2^\ast=0$
should violently oscillate with an accumulation point, infinitely many solutions
might be possible. The more realistic and pertinent 
tangential case of $g_2^{\ast\ast}=0$ will be encountered again later
and discussed then.

Because no a priori insight into the expected topologies 
of the $g_1=0$ and $g_2=0$ loci on the $P_{\text{c}} - T_{\text{c}}$ plane
is possible, i.e. no direct answer to the question of global uniqueness
appeared to be accessible, K\"ahler resolved to concentrate first on 
the simpler, \emph{local uniqueness }question.

Starting out with assuming that an equilibrium solution, i.e. $g_1=0$ and 
$g_2=0$ exists, K\"ahler inquired the properties of the fitting conditions 
in the direct neighborhood of such an equilibrium solution. The linear
approximation of the functions $g_1=0$ and $g_2=0$ at the found solution
leads to the condition
\begin{equation}
	\mathbb{J}_g \cdot \left( 
			              \begin{array}{c}
                             \delta P_{\text{c}}\\
	                         \delta T_{\text{c}}\\
                           \end{array} 
                       \right) = 0
    \label{eq:localuniquness_hom}
\end{equation}
for an additional solution displaced by 
$\left(\delta P_{\text{c}} \,\,\delta T_{\text{c}} \right)^T$ on 
the $P_{\text{c}} - T_{\text{c}}$ plane with $\wp$ remaining 
unchanged and with 
\begin{equation*}
	\mathbb{J}_g \doteq \left( 
			              \begin{array}{cc}
                             \partial_{P_{\text{c}}}\,g_1 & \partial_{T_{\text{c}}}\,g_1 \\
                             \partial_{P_{\text{c}}}\,g_2 & \partial_{T_{\text{c}}}\,g_2 \\  
                           \end{array} 
                       \right) \,, 
\end{equation*}
being the Jacobian of the fitting conditions $g$. For $\det \mathbb{J}_g \neq 0$, 
the equilibrium stellar-structure problem is \emph{locally unique} because only 
$\left(\delta P_{\text{c}} \,\,\delta T_{\text{c}} \right)^T = \vec{0}$
obtains; i.e. no other FE star model with the same set of parameters 
$\wp$ can be found in the neighborhood of an already obtained one. 

The case of $\det \mathbb{J}_g = 0$, on the other hand, allows for a nonvanishing
$\left(\delta P_{\text{c}} \,\,\delta T_{\text{c}} \right)^T$ vector, i.e.
for another solution to the FE equations, in an essentially 
arbitrarily close neighborhood of an already existing one for the same 
choice of parameters $\wp$. Hence,  $\det \mathbb{J}_g = 0$ amounts 
to a violation of the classical VR theorem formulation
(such as illustrated in intersections of $g_1=0$ with 
$g_2^{\ast}=0$ or $g_2^{\ast\ast}=0$ in Fig.~\ref{fig:IntersectionOptions}).
An intersection case such as $g_1=0$ and $g_2^{\ast}=0$, with its 
accumulation point, seems unlikely to be physically realized in naturalistic 
stellar astrophysics but mathematically it remains a viable option. 
Contacts such as encountered in $g_1=0$ and $g_2^{\ast\ast}=0$, with
common tangents, are involved whenever a star changes from secular 
stability to instability or vice versa. 

An appendix of the \citet{Kaehler1972} article was devoted to the 
proof that the vanishing of $\det \mathbb{J}_g$ was equivalent 
to the vanishing of the Henyey-determinant; i.e. the determinant 
of the finite-difference system that constitutes the numerical 
representation of the model
stars.\sidenote[][-0.0cm]{Henyey's scheme to solve the quasi-static stellar
  evolution problem applies essentially the Thomas-Algorithm to solve a large
  block bi-diagonal matrix. The Henyey-{\it determinant }can be easily
  obtained from computing the determinant of the last block matrix
  of during forward elimination. N.B.~The magnitude of the
  Henyey-determinant depends on the gridding of the star model; the sign
  of the determinant, however, is invariant under regridding.}

\newthought{Variations of the parameter }$\wp$ is a method to compute linear series
of stellar models in FE. Presuming again that
an equilibrium stellar model had been found for a choice of $\wp$, one asks
if another equilibrium solution, displaced on the 
$P_{\text{c}} - T_{\text{c}}$ plane by 
$\left(\delta P_{\text{c}} \,\,\delta T_{\text{c}} \right)^T$,  
can be found upon varying the parameter by $\delta\wp$.
The resulting linear system of equations becomes 
\begin{equation}
	\mathbb{J}_g \cdot \left( 
			              \begin{array}{c}
                             \delta P_{\text{c}}\\
	                         \delta T_{\text{c}}\\
                           \end{array} 
                       \right) = - \delta \wp
                       \left( 
			              \begin{array}{c}
                             \partial_{\wp} g_1 \\
	                         \partial_{\wp} g_2 \\
                           \end{array} 
                       \right)\,,
\label{eq:PrmtrVariation} 
\end{equation}
which is just an inhomogeneous form of eq.~(\ref{eq:localuniquness_hom}).
Equation~(\ref{eq:PrmtrVariation}) admits of an unique
solution $\left(\delta P_{\text{c}} \,\,\delta T_{\text{c}} \right)^T$
upon a parameter-change $\delta \wp$ if $\det \mathbb{J}_g \neq 0$. 
Hence, as long as an equilibrium model is unique, a unique model continuation 
can be constructed upon variation of the control parameter $\wp$.
Such a parameter-$\wp$ viewpoint can be adopted in the study of properties 
of say hydrogen or helium main sequences with the stellar mass as the control
parameter. A suitable parameterization of the chemical composition, 
which be prescribed in $\wp$, can be thought of as an emulation of the 
evolution of stars in FE. Such equilibrium models remain 
locally unique as long as $\det \mathbb{J}_g$ does not vanish.
In the inhomogeneous case of eq.~(\ref{eq:PrmtrVariation}) this
means that a unique solution obtains if the rank of the augmented system is
equal to $\mathrm{rank} \,\mathbb{J}_g$. If, on the other hand, 
$\det \mathbb{J}_g = 0$, the number of local solutions depends again on 
the rank of the augmented system. The case of no solution signifies that 
the linear series reached a termination point. 
Alternatively, a branching of solutions, such as at the onset of   
the Sch\"onberg-Chandrasekhar instability or blue-looping  core helium-burning
intermediate-mass stars, as illustrated in Fig.\ref{fig:Fig7b}, can be encountered. 

\newthought{Quasi-hydrostatic equilibrium (QHE) models} were 
the next logical step to take in the study of local uniqueness properties 
of stellar-evolution models. The paper of \citet{Kaehler1974} followed 
quite closely the approach used in \citet{Kaehler1972},
but now incorporating the thermal imbalance term $ \Diff_t s$. 
To get a better handle of the equations, the thermodynamic basis defined by 
the total pressure and the specific entropy, $\left\{ P,s \right\}$, 
was found to be more appropriate. 
The formal structure of the QHE stellar-structure equations becomes then:
\begin{alignat}{2}
\label{eq:QHE_radius}
  \partial_m r & = f_{1} \left(r,P,s \right)           & \quad\left(\text{from I.2}\right)\,, \\
\label{eq:QHE_pressure}  
  \partial_m P & = f_{2} \left(m,r    \right)          & \quad\left(\text{from I.6}\right)\,, \\
\label{eq:QHE_temp}
  \partial_m s & = f_{3} \left(m,r,L,P,s \right)       & \quad\left(\text{from I.10}\right)
\,, \\
\label{eq:QHE_lum}
  \partial_m L & = f_{4} \left(P,s, \partial_t s \right)  & \quad\left(\text{from I.8}\right)\,.    
\end{alignat}
\marginnote[-2.5cm]{Equation numbers prepended with '$I$' refer to those 
used in the first installment on the VR theorem \citep{Gautschy2015}.}
The resulting system of equations is hence no longer a 
BVP of ordinary differential equation. The problem turns into an 
initial-value~--~boundary-value problem which requires the theory of
partial differential equations. However, the mathematical theory of PDEs is not 
general enough to put forth helpful statements for the types of equations
popping up in the description of stellar structure/evolution problems.
The situation can be improved when adopting a \emph{prescribed entropy profile}, 
$s(m)$, for the stellar model about which the local (linear) analysis is  
to be performed. In this case the 
equations~(\ref{eq:QHE_radius})~--~(\ref{eq:QHE_temp}) 
fall back onto a BVP of ordinary differential equations and the same 
line of arguments applies as in the FE case earlier on.

Analogously to the FE case, in-out and out-in integrations are performed, 
which are matched at some fitting mass $m_F$. With a prescribed
entropy profile, equations~(\ref{eq:QHE_radius}) and (\ref{eq:QHE_pressure}), 
i.e. the star's mechanical equations decouple from the thermo-energetic part 
(eqs.~(\ref{eq:QHE_temp}) and (\ref{eq:QHE_lum})). 
The interior, mechanical solutions 
can be parameterized via the central pressure, $P_\mathrm{c}$, alone. 
Once again, exterior solutions, i.e. out-in integrations are 
performed to $m_F$ where the solutions are matched. 
In the QHE case, $r_\mathrm{I}(P_\mathrm{c}) = r_\mathrm{O}(R)$, is used to find 
a relation $R= R(P_\mathrm{c})$ so that, eventually, 
the fitting condition, analogous to eq.~(\ref{eq:g_1}) in the FE case, 
can be determined:
\begin{equation}
	g(P_\mathrm{c}) = P_\mathrm{O}(R(P_\mathrm{c})) - P_\mathrm{I}(P_\mathrm{c}) \,.
\end{equation} 
Note, that once the mechanical structure is determined, the thermal one follows:
The prescribed entropy profile allows to compute the luminosity profile in
eq.~(\ref{eq:QHE_temp}) and with its spatial derivative, the $\Diff_t s$ profile
can be obtained via eq.~(\ref{eq:QHE_lum}):
\begin{equation}
\Diff_t s = \frac{1}{T}\left[ \varepsilon(P,s) - \Diff_m L \right] \,,
\label{eq:Dts}
\end{equation}
so that the temporal evolution of a star's entropy distribution is determined.

Notice furthermore that the prescription of the entropy profile reduces the number of 
fit-equations to one. Hence, the fitting procedure, and the linear perturbation
analysis reduces to the study of scalar equations rather than to matrix properties
as in the FE case.

Linearizing about a solution, varying $P_\mathrm{c}$, keeping fixed the mass, 
the chemical profile, as well as $s(m)$, i.e. with all of $\wp$ unchanged, yields
\begin{equation}
 \diff_{P_\mathrm{c}}\,g \cdot \delta P_\mathrm{c} = 0 \,.
\end{equation} 
For $\diff_{P_\mathrm{c}}\,g \neq 0$, the only solution is  
$\delta P_\mathrm{c} = 0$; i.e. no neighboring solution can be found so
that the found QHE solution is \emph{locally unique}. On the other hand, 
as already seen in the FE case, for $\diff_{P_\mathrm{c}}\,g = 0$, 
the solution is not locally unique and possibly multiple solutions exist 
in an arbitrary neighborhood of the already obtained QHE solution. 

In case of a variation of parameters, the case of a modified entropy profile
is now of particular interest (the rest, variation of mass and chemical profile
is analogous to the FE situation). Assume that we write:
\begin{equation*}
	s(m,\wp) = s_0(m,\wp_0) + f(m)\cdot\delta\wp \,,
\end{equation*}
with $\delta \wp \in \mathbb{R}$ and $f(m)$ an arbitrary, 
continuous function of mass alone.
Linearization about a solution for $s_0(m,\wp_0)$:
\begin{equation}
\label{eq:QHE_VarParms}
 \partial_{P_\mathrm{c}}\,g \cdot \delta P_\mathrm{c} 
+  
 \partial_{\wp}\,g \cdot \delta \wp
 = 0 \,.
\end{equation}

Starting from a locally unique QHE solution with $s_0(m)$, 
eq.~(\ref{eq:QHE_VarParms}) yields
\emph{one } neighboring solution for a slightly changed entropy profile if
$\partial_{P_\mathrm{c}}\,g \neq 0$ and  $\partial_{\wp}\,g \neq 0$.
This means that a model sequence with varying entropy profile, say, 
according 
to\marginnote{Prescribing the temporal evolution of the entropy profile
	          in an Euler-type numerical scheme as 
	          suggested on the left is theoretically 
	          fine. In practice, however, it turns out that
	          the \emph{explicit }integration of thermal 
	          variables in model stars is frequently numerically unstable
	          and must be avoided \citep[cf.][]{sugimoto70}.
	           }
\begin{equation*}
	s(m,\wp;t) = s_0(m,\wp;t_0) + \Delta t \cdot \left. \Diff_t s \right|_{t_0} \,,
\end{equation*}
with $\Diff_t s$ computed from eq.~(\ref{eq:Dts}), can be continued
continuously as long as the QHE model is locally stable.
If a model violates local uniqueness $(\partial_{P_\mathrm{c}}\,g = 0)$ then
$\partial_{\wp}\,g \neq 0$ suppresses any local solutions, but  
$\partial_{\wp}\,g = 0$ allows for multiple ones in the vicinity of $s_0$.

\smallskip
\newthought{Stellar stability}, at least in the linear approximation, 
can be related to local uniqueness, this aspect is also computationally interesting.
In QHE modeling, with its thermal imbalance term (but neglected acceleration), 
assume a time independent FE solution with
physical quantities $\hat{y}_i$; perturb this solution by some $\delta y_i$: 
\begin{equation*}
  y_i(m,t) = \hat{y}_i + \delta y_i \,, \quad \mathrm{adopting} \quad
  \delta y_i = \upsilon_i(m) \cdot \exp(\sigma t) \,.
\end{equation*}
Together with appropriate boundary conditions, a boundary eigenvalue
problem results for the perturbations of the equilibrium 
state, $\upsilon_i$, become spatial eigenfunctions for a discrete set of   
eigenvalues~$\sigma$.\sidenote[][-2cm]{The physical nature of the time 
dependence entering the eigenvalue problem leads to it being referred 
to as thermal or secular stability problem. In contrast to the 
adiabatic pulsation problem, the secular eigenproblem is not Hermitian 
and lacks therefore any elegant properties of the eigensolutions. 
Most importantly, its eigenvalues $\sigma$ can be complex.} 
For simplicity, assume first that the secular eigenvalues are all real.
If an initially secularly stable model sequence is computed 
then all the associated eigenvalues $\sigma < 0$. If later, during 
the star's evolution secular instability sets in, the eigenfrequency of the 
respective secular mode goes positive. Hence, an epoch is 
encountered when $\sigma=0$:
\begin{equation*}
  y_i(m,t) = \hat{y}_i + \upsilon_i(m) \,,
\end{equation*}
which is time independent, and which therefore violates local uniqueness 
with $\det \mathbb{J}_g = 0$ at this particular epoch: It was mentioned
further up that \citep{Kaehler1972} pointed out the connection of 
$\det \mathbb{J}_g = 0$ and zeros of Henyey-determinant.
At the very me time, also Paczy\'{n}ski was active in the same field; 
his paper \citep{Paczynski1972a} reached the editorial
office of Acta Astronomica on March 15th, 1972 (K\"ahler's paper 
arrived at A\&A's editorial office on the 3rd of March, 1972).
In the same manner as K\"ahler, Paczy\'{n}ski elaborated on the
Henyey determinant, its zeros (which coincide with the zeros of the
Schwarzschild
determinant\sidenote[][]{The Schwarzschild determinant is characterized by
  the fitting conditions of trial solutions as obtained from the IVP
  integrations once from the surface and once from the center to a
  prescribed fitting point, respectively.})
and the connection of the roots with the secular stability of a stellar model.
In particular, the passage of stars through the
Sch\"onberg-Chandrasekhar instability~--~if they are approximated by a
linear series of full-equilibrium models~--~leads to a phase of three
equilibrium solutions for the same set of stellar parameters, with one
branch being secularly 
unstable.\sidenote[][]{The effect had been encountered and discussed in
	\citet{Schwarzschild1965} and \citet{Gabriel1967}.} 
The same conclusion was also arrived at
within the methodical framework of \citet{Kaehler1972}.

In the case of \emph{oscillatory} secular modes, the 
Henyey-/Schwarzschild-determinant does not vanish anymore so that the onset
of secular instabilities via oscillatory modes cannot be tracked by means of
monitoring the sign of the Henyey determinant. 
However, for complex eigensolutions with not
too large oscillation frequencies, their change of stability is frequently
noticeable in the determinant by its dipping through a local minimum as the 
model sequence progresses.

Irrespective of the path to instability of $\sigma$ on the complex plane, 
it became clear that 
$\mathrm{sign}\left( \det \mathbb{J}_g\right) = +1$ is a necessary 
requirement for the model to be thermally stable; 
$\mathrm{sign}\left( \det \mathbb{J}_g\right) = -1$, on the other hand, 
is a sufficient condition for instability.

\section*{The global perspective}
Local analyses boil down to a linear expansions of the curves $g_1 = 0$ and
$g_2=0$ about solutions $s_0$; they are, however, only of limited use to understand 
the full solution topology. The approach breaks down if, for a chosen
$\wp$, multiple solutions exist at finite distance of each other on the 
$P_{\text{c}} - T_{\text{c}}$ plane or if a solution is not a regular one, 
i.e. if  $\det \mathbb{J}_g = 0$.
Applying heavier mathematical machinery than usual in astrophysics, 
\citet{Kaehler1975} drew from the field of algebraic geometry and found a way to 
tackle even such non-local problems. The resulting abstract, 
mathematically convoluted paper is though not for the formally 
fainthearted.\sidenote[][-2cm]{Helmuth K\"ahler is the son of the eminent
German mathematician Erich K\"ahler who is known, among other things, 
for contributions to algebraic geometry. Hence, it is well conceivable that 
young K\"ahler got good advice from within his family circle on mathematics 
that usually lies beyond the astrophysicists' horizon .} 
To that effect, the paper had essentially no impact on the field
(with only two citations from outside of the Hamburg group). Up to the 1994 
edition of the Kippenhahn \& Weigert textbook,  its chapter 12 contained  
a digested synopsis of the contents of the 
\citet{Kaehler1975} study, attempting to make it more palatable to 
the students of the stars. 

To retrace the path to K\"ahler's findings, introduce the vector field 
${\bf g} = \left( g_1, g_2 \right)$ on the  $P_{\text{c}} - T_{\text{c}}$ 
plane.\sidenote[][-0.5cm]{The characteristic plane can also be defined by other
suitable stellar-physical quantities, such as $R_\ast, L_\ast$, which 
then define a coordinate system homeomorphic to the HR plane.}
To study the nonlocal solution properties, consider first the loci
$g_1 = 0$ and $g_2 = 0$ for a given set of $\wp$. In the neighborhood of a
solution $s_0$ of the stellar structure problem, $g_1$ and $g_2$ are both 
assumed to be representable by algebraic equations; this can be accomplished
if  $g_1$ and $g_2$ are expanded in higher-order Taylor series about $s_0$.
B{\'e}zout's theorem allows to count the \emph{number of intersections }of 
algebraic curves on the basis of their degrees. 
K\"ahler referred to this intersection number as 
\emph{multiplicity, }$m$, of the joint roots of $g_1 = 0$ and $g_2 = 0$.
The number $m$ gives the maximum number of solutions to the
stellar structure problem at  $P_{\text{c}} - T_{\text{c}}$.
Regular solutions, with $\det \mathbb{J}_g \neq 0$, have $m=1$. 
In case of multiple solutions, i.e. $\det \mathbb{J}_g = 0$ violating
local uniqueness, $m > 1$. A double solution, for example, 
such as $g_1$ and $g_2^{\ast\ast}$ in Fig.~4, counts as $m=2$. 
 
In a next step, K\"ahler studied the character of solutions $s$, making
use of the nature of the singularities ${\bf g} = \left( 0, 0 \right)$
of the vector field ${\bf g}$. For regular singularities, i.e. for 
locally unique solutions $s$ with $m=1$, he resorted to the  
$\mathrm{sign}\left( \det \mathbb{J}_g\right)$ and baptized this
quantity the \emph{charge }of a model star. For an $m=1$ solution, $c$ is
either $+1$ or $-1$. For degenerate solutions, $c=0$ in case of even $m$, 
and $c=\pm 1$ for odd $m$. 
Setting up an admissible closed path $B$ on the $P_{\text{c}} - T_{\text{c}}$
plane (with no singularities on the locus) the Poincar\'e index $C$ 
of ${\bf g}$ can be computed. This quantity $C$ is made up of the sum 
over the charges, $c_i$, of the $B$-enclosing singularities (solutions)
$s_i$:\sidenote{ 
	This is reminiscent of the computation of the total electric charge contained 
	in a spatial region by integrating over the surface of the volume 
	enclosing the point charges, and explains why K\"ahler 
	referred to $c$ as the \emph{charge }of a stellar model.   
	}
\begin{equation*}
C = \sum_i c(s_i(\wp)) \,.
\end{equation*}
By just evaluating a path integral, information can be gained about 
the charges captured by the closed path. 
Along the same line, the total number of solutions, $N = \sum m_i$, 
is just the sum of the solutions $s_i(\wp)$ enclosed by $B$, accounting
correctly for respective multiplicities in the sense algebraic geometry. 

If $\wp$ varies continuously (such as in linear series of stellar models), 
the total charge $C$ is found to be conserved. The total number of solutions, 
on the other hand, is either constant or changes by even numbers. In other
words, varying the control parameter of a linear series lets new solutions 
dis-/appear in pairs. 

Since both, $c$ and $m$, attain integer numbers only and since both numbers
admit of conservation properties or at least follow some well defined selection
properties, K\"ahler considered them as the \emph{quantum numbers }of 
a model star. 

In terms of $c$ and $m$, K\"ahler concluded that a necessary 
condition for unique stable solutions to the stellar structure problem 
is: $m=c=+1$. Making use of the conservation properties of these 
stellar quantum numbers under $\wp$ variation, the claim is then
that always at least one stable stellar model exists (amounting to an
existence theorem); additional solutions might pop 
up as $\wp$ changes. These solutions appear in pairs, always with 
a stable one and one being thermally unstable 
(hence, uniqueness prevails even locally).    

\begin{figure} 
	\begin{center}
	  \includegraphics[width=0.85\textwidth]{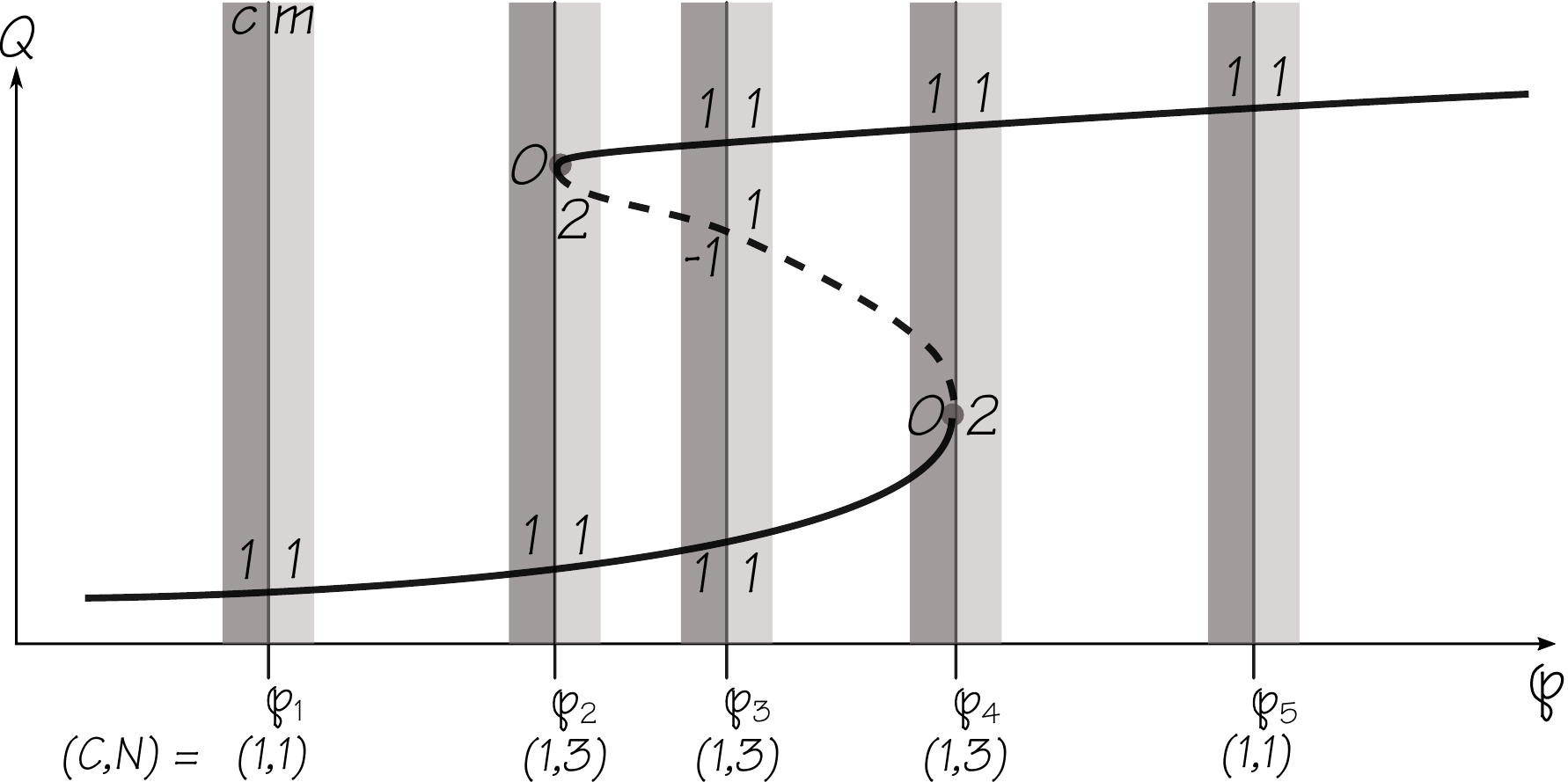}
	\end{center} 
	\caption{Variation of the \emph{quantum numbers }$c$ and $m$ along the
		     a linear series (with control parameter $\wp$) with two 
		     turning points. The quantity $Q$ could be effective temperature
		     such as in the case illustrated in Fig.~\ref{fig:Fig7b}. 
	        } \label{fig:GlobalStmts}
\end{figure}

Figure~\ref{fig:GlobalStmts} illustrates the variation of the \emph{quantum numbers}
$c$ and $m$, as well as of $C$ and $N$ for a model sequence as presented in
Fig.~\ref{fig:Fig7b} wherein the star's core mass is the control 
parameter $\wp$ and $Q$ measures the model star's effective temperature. 
The model sequence ``evolves'' from $\wp_1$ to $\wp_5$. 
Epochs at $\wp_4$ and $\wp_2$ are turning points. The solutions between the
turning points constitute the \emph{branches} of the model sequence. 
Evidently, according to the counting rules of K\"ahler, $C=\sum c$ 
is always $+1$, and $N=\sum m$ is $1$ early on, for epochs before $\wp_2$ and for
epochs after $\wp_4$. In between, $N=3$, being unity plus  
an even number. This behavior is canonical for solution loci with turning points.
If $Q$ of Fig.~\ref{fig:GlobalStmts} would be suitably flipped upside 
down, the result would be reminiscent of a sufficiently massive star's core radius
as a function of its growing core mass when the respective equilibrium-star model
passed through the Sch\"onberg-Chandrasekhar instability. 

\begin{marginfigure}[-2.8cm] 
	\begin{center}
	\includegraphics[width=0.97\textwidth]{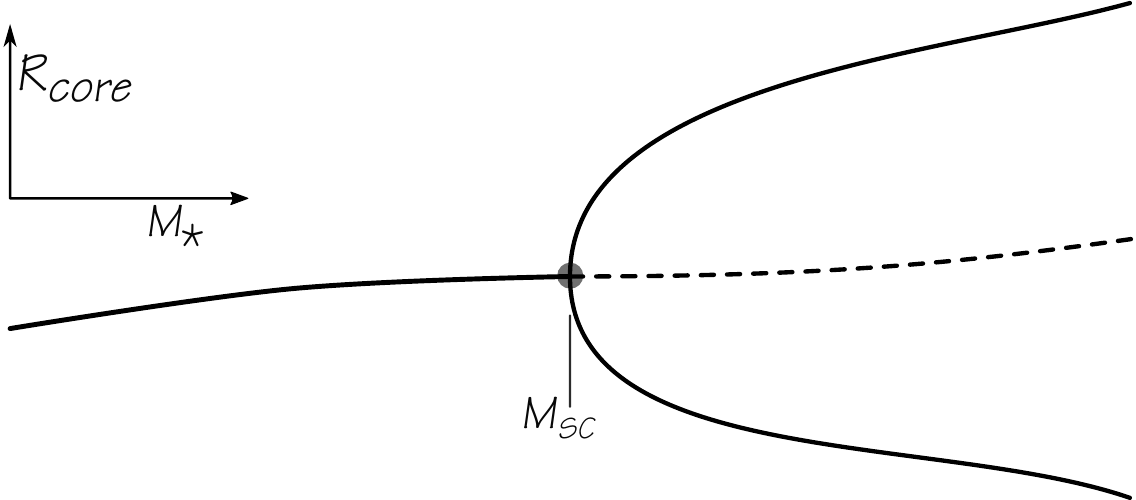} 
	\end{center}
     \caption{The onset of the Sch\"onberg-Chandrasekhar instability 
     	       as seen in a stellar mass, $M_\ast$, core-radius, 
     	       $R_\mathrm{core}$, diagram of equilibrium models.      
             } \label{fig:Bifurcation}
\end{marginfigure}

Other solution topologies that show up in stellar modeling are 
\emph{termination points }(as encountered if \emph{no} model
star is possible below/above some critical value of $\wp$) and
\emph{bifurcations }that appear at the onset of the Sch\"onberg-Chandrasekhar
instability (at $M_\mathrm{SC}$) in the \emph{total mass}~--~core-radius diagram.
The illustrative example of Fig.~\ref{fig:Bifurcation} is made up 
of equilibrium models with $m=3$ and $c=+1$ \citep[e.g.][]{Paczynski1972a}. 

\smallskip
Rather than focusing on the behavior of trial integrations of the
differential equations of stellar structure, the analysis of the properties
of the differential operators constitutes an alternative approach to 
tackle the VR problem. This latter approach 
calls for tools and results in nonlinear functional analysis.  
At about the same time as \citet{Kaehler1975} dealt with algebraic geometry, 
\citet{Perdang1975} traveled the path of functional analysis 
and published a rather formal mathematical treatise discussing existence and 
uniqueness of solutions.\sidenote[][-0.0cm]{Perdang submitted his 
paper for publication about half a year before K\"ahler. 
According to the acknowledgments in both papers, the two authors seem 
to have discussed the overlapping topics, possibly when their 
paths crossed at Columbia University where both of them spent some 
time in the Department of Astronomy. }
For that approach to be tractable at all, the model stars had to be 
sufficiently simplified: The stars were
assumed to be in full equilibrium and completely radiative. 
First, Perdang recast the stellar structure equations into two higher-order 
ones, a mechanical and a thermal one. The direct coupling of the two 
components occurred via the matter density as well as the nuclear
energy generation rate and the thermal conductivity, which are functions 
of pressure and temperature (i.e. the thermodynamic basis variables, 
which are also and the key variables of the mechanical and the thermal 
equations). At the surface, finite pressure and finite temperature were assumed 
to prevail. After transforming the variables into dimensionless form 
and some heavier mathematical machinery of function spaces, norms, 
and products as it is customary in functional analysis, 
Perdang rewrote the two equations into integral form
and morphed them into a nonlinear integral operator to which Perdang
referred as the \emph{stellar structure operator}. The stellar structure
problem was thereby transformed into an eigenvalue problem for the stellar radius.

Under rather general restrictions (continuity with respect to the
thermodynamical basis, single-valuedness, first and second partial 
derivatives do exist) imposed on the constitutive relations~--~equation 
of state, opacities, nuclear energy generation rates~--~Perdang found 
that at least one solution $R_\ast, P(r), T(r)$ exists if $R_\ast$ stayed 
smaller than some prescribed value $\cal{R}$; such a solution can even 
be found to be unique if $R_\ast$ is smaller than a more restrictive
bound $\cal{R}'$.

Perdang scrutinized his approach on simple model systems such as polytropes
and isothermal, self-gravitating gas spheres. Because the radius plays the
role of an eigenvalue in the integral-equation formulation, he used
the stellar radius as a control parameter in linear model series. 
For the simple case of self-gravitating isothermal gas spheres,
the properties of Bonnor-Ebert spheres, which are known
to be unable to achieve an equilibrium state for too high external pressures,
were recovered. From the point of view of linear series, this means 
that sequences of isothermal gas spheres, embedded in a pressure-exerting 
external medium, pass through a turning point so that above a critical 
radius no equilibrium configurations can persist anymore. 
Actually, Perdang pointed out that his 
existence and uniqueness arguments obtain only in case of radius being the 
control parameter of the series but not if mass were adopted. 

Despite his efforts, also Perdang failed to generate momentum 
to motivate further work along his line to probe the fundamental 
properties of stars. The paper garnered only two citations through the years. 
Ultimately, the model stars that were accessible to Perdang's approach 
are likely to have been too abstracted and the math itself too abstract to 
entice any students of the stars.   

\section*{The state of the affair} 
The paper of 
\citet{Kaehler1975}\sidenote{In a conference paper, \citet{Kaehler1978} tried 
to clarify and digest somewhat his main contribution from 1975.}  
and the singular one by
\citet{Perdang1975} marked the endpoints of the few years 
of intensive research on existence and uniqueness of stellar models in 
general and on the VR theorem in particular.

Admittedly, the results on the VR theorem do not directly affect 
the practice of stellar modeling, in particular not if data-oriented 
computation is the focus. Even if multiple solutions to the 
stellar-structure/-evolution equations 
are possible, the history of a star imposes the ensuing direction of evolution. 
This starts with the condensation of the protostars 
out of the low-density, more or less homogeneous interstellar material, which
governs the evolution from low to high densities and temperatures, and 
from simple to complex chemical composition and spatial structure. 
Structural possibilities that might also be possible, at least mathematically,
are therefore automatically excluded.  

All in all, star models are locally unique except at epochs where a 
model's stability changes. At these critical points
uniqueness is lost. Physically, the situation can be usually
saved by then adopting a more comprising description of the problem 
(as going from FE to QHE, or going from QHE to dynamical evolution, for example).
From the viewpoint of global properties of the solution manifold, 
the existence of at least one stable equilibrium solution is claimed to be
guaranteed with reasonably weak restrictions only on the model properties 
\citep{Kaehler1975}. When applying the result of the conservation of the 
charge of a model star upon continuously varying components of 
its $\wp$ vector, it is reasonable to conclude that at least one 
stable model can be found on the HR plane (or equivalently on a plane 
homeophorphic to it). On the other hand, uniqueness is not ensured.  
As before, in the local context, a star's history will nonetheless select 
a particular set of solutions, so that additional branches are not accessible 
to a model star and can therefore usually be neglected in the analyses. 
 
Even though important results elucidating the connection between evolution 
and stellar stability were obtained in the 1970s, the model stars that were
studied then had to be necessarily simple. 
Mathematical analyses of the more complex systems of equations that 
describe say dynamical stars are not yet in (see Appendix A). 
Substantial changes of the results as obtained up to now would, however, be 
surprising. Nonetheless, it would feel good to know that the foundations
and the computations are on solid ground. 
Last but not least, the Universe itself has proven 
to be a wilder place than humans' imagination. 
So we can~--~particularly now as the next Cosmic data deluge builds 
up~--~rest confident that if seemingly weird stellar configurations 
are realized somewhere in the Universe, sooner or later we are 
going to stumble over them and we will, as usual in astronomy, likely 
come to grips with them \emph{post festum }by means of reverse engineering. 
 
In any case, if stellar astrophysics aspires to be more than a branch 
of celestial engineering one could do worse than to sit back time and again 
and contemplate qualitatively fundamental questions framing the field 
of research. Here, these questions concerned the solution properties of the basic 
equations describing the macroscopic structure of the stars.  

\section*{Appendix A : The nature of the equations - II}
\label{sec:ExistenceUniqueness}
Evolving model stars that yield helpful results to astronomers is
traditionally, for computational reasons, a one-dimensional
enterprise.  Resorting to suitable parameterizations to model
deviations from radial symmetry due to rotation and/or tidal effects
in close binaries even allow to continue along the one-dimensional
path of simulating stars.

At full glory, though, stellar structure and evolution is a
three-dimensional problem. The computational methods and the computer
power are, however, far from ready to embark on full-scale stellar
evolution simulations covering nuclear timescales in three dimensions 
in a foreseeable future. Nonetheless, the
properties of the underlying equations can be studied, and this is
where \emph{mathematical fluid dynamics }enters the stage.

\smallskip

Two main avenues of mathematical modeling, independent of the
dimensionality of the spatial description, are encountered:

\begin{enumerate} 
\item Star models in FE constitute a
  boundary-value problem of an ODE system; evolution can be approximated
  by a prescribed chemical profile (see linear series) at the discretion
  of the modeler.
\item Star models with 'built-in' temporal evolution change the cha\-rac\-ter
      of the mathematical problem to an initial~--~boundary-value 
      problem of a PDE system not belonging to any standard class. 
\end{enumerate}

Mathematically speaking, the most elementary stellar \emph{structure} 
problem (say in FE) is a two-sided boundary-value problem of Euler-Poisson 
type. If energy transport through the stellar matter is modeled via 
photon diffusion (eq.~I9) or some sort of a mean-field convection ansatz,
the problem acquires a \emph{diffusion-type }(hyperbolic) contribution.
Last but not least, nuclear energy sources responsible for the
long-term energy supply of the star add further equations
of \emph{reaction-diffusion type }(eq.~I11) to the problem.

The paper of \citet{Makino1986} was an early contribution to the
discussion of the existence of solutions of the three-dimensional
Euler-Poisson system, adopting a barotropic equation of state and
neglecting the effect of radiation. The compact support of the gas
ball (i.e. the spatial confinement of the stellar matter) posed
mathematical problems which Makino set out to solve. Eventually, he
managed to formulate sufficient conditions on initial data for
short-time existence of solutions to the Euler-Poisson system, 
which applied to polytropic indices $n > 2$.

Even though stars are usually considered low-viscosity
fluid-dynamical environments, turbulent convection, winds, high
radiation energy densities, shocks and other nonlinear fluid-dynamical
processes require the Euler equation to be replaced by its
Navier-Stokes brethren.
Because the Millennium-Prize money for the \emph{incompressible}
multidimensional Navier-Stokes problem seems not to have been
disbursed yet, the jury is still out on proofs of the existence 
of regular solutions given smooth initial data even on a simpler
Navier-Stokes~--~type problem than what is needed for stars. 

The mathematicians' closest approach to an astrophysically pertinent
radiating, self-gravitating gas blob was achieved in the paper of
\citet{Ducomet1996}. That paper outlines the equations in their the
most general form. In all approaches, though, the equation of state is
that of an ideal gas with
radiation\sidenote{The effect of radiation is accounted for by adding
 a Fourier-type diffusion equation for heat to the system. In other
words, such models describe purely radiative model stars, such as dealt
with by \citet{Perdang1975}.}
and the thermal conductivity in
the heat-diffusion equation is considered to be fixed. The resulting 
full set of equations had to be simplified drastically before local and
global existence theorems could be proved.
 
For short times, \citet{Secchi1991, Secchi1990} proved existence and
uniqueness theorems for Navier-Stokes~--~Poisson~--~Fourier systems.
\citet{Ducomet2004} returned to the problem stated by Ducomet in 1996
and put forth an existence theory for three-dimensional weak solutions
that hold, in contrast to Secchi's results, for arbitrary time
intervals.

Much activity in mathematical fluid-dynamics went also in studying
subproblems: For example, \citet{Deng2006} studied stationary
solutions of three-dimensional isentropic Euler-Poisson systems;
\citet{Xie2012} extended the analysis to non-isentropic equations of
state. The stationary velocity fields that were accounted for can be
thought of as to deal with rotating stars. Uniqueness of solutions of
simplified spherically symmetric Navier-Stokes systems with conduction
could be shown by \citet{Umehara2007}.

In all the studies of the type mentioned just above, it is evident
that the mathematical problems associated with the full
set of equations that describe radiating, self-gravitating gas balls are so 
formidable that the equations need to be stripped down dramatically
to be mathematically tractable. In the end, it is always difficult for 
outsiders to figure out if the resulting mathematical formulation is still 
germane to stellar physics.   

\section*{Appendix B: Computational exercitia}
To put some flesh on the dry, abstract bones contemplated up to now, 
pertinent properties of one-dimensional, canonical {\tt MESA}-computed
$5 \msol$ and $15 \msol$ star models are presented in the
following.

\subsection{B.1. A $5\,\msol$ star from ZAMS to core He-burning}
\label{sec:FiveMsol}
\citet{Murai1974} pointed to the functional behavior of the opacity in
the deep envelopes as the reason for multiple solutions, i.e. of
secular instabilities in intermediate-mass model stars.  Hence,
in contrast to models with canoncial $(\rho,T)$-dependent microphysics,
stellar models resorting to constant electron-scattering~--~only opacity
should therefore lack the hook in the fitting curves on the $\log P$~--~$\log T$
plane and hence multiple solutions should thereby be suppressed.

\begin{figure} 
	\begin{center}
	  \includegraphics[width=0.85\textwidth]{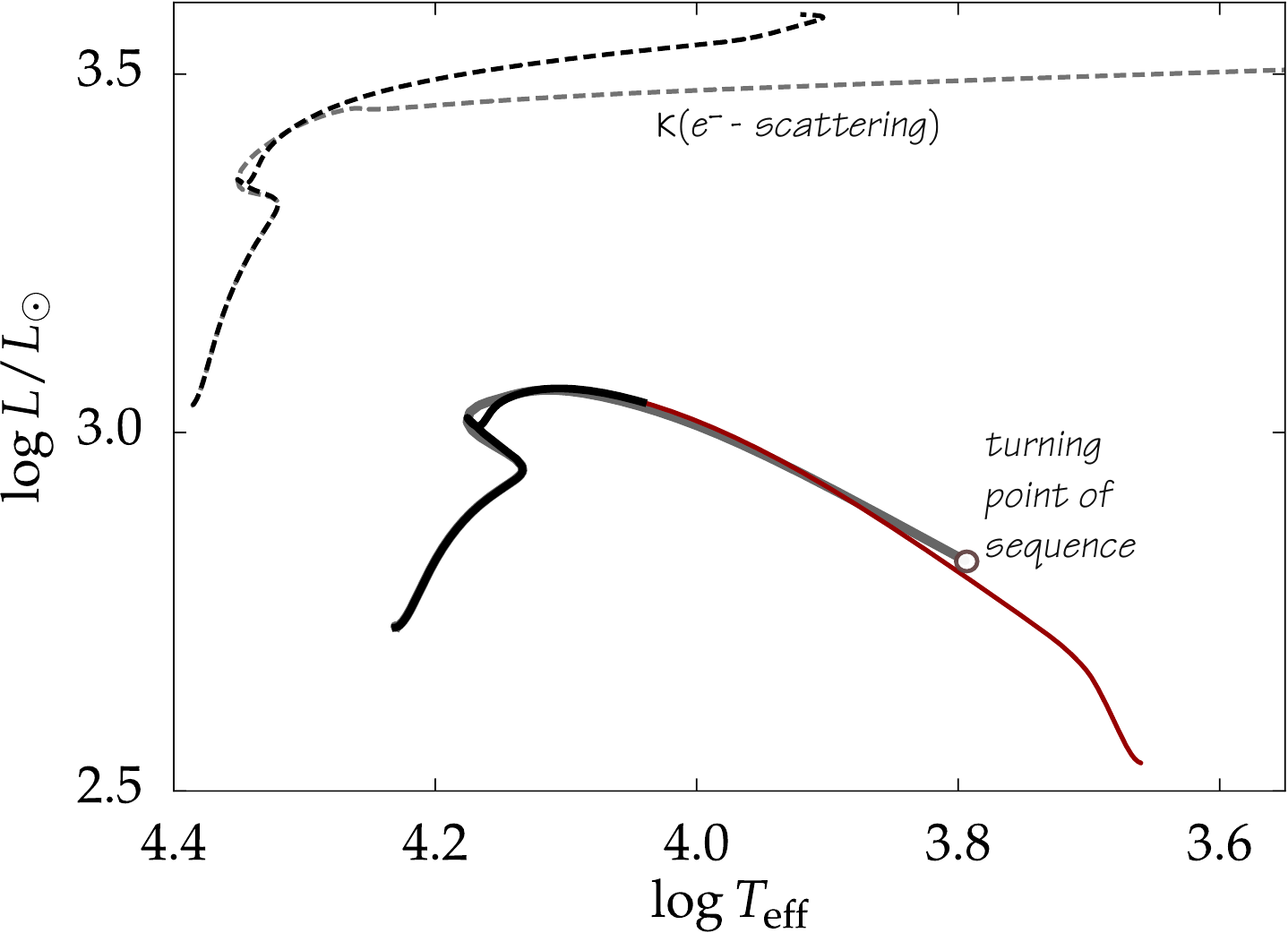}
	\end{center} 
	\caption{Exemplary evolutionary tracks of $5 \msol$ stars with $X=0.7,$ 
	      	 $Z=0.02$. 
	         Full lines: Loci computed with OPAL opacities. 
	         Dashed lines: Tracks of models with pure
                 ${\mathrm{e}^-}$-scattering opacity. 
	         Black lines: Quasi-static evolution: Grey lines: 
                 Full equilibrium models.
	        } \label{fig:m050z02}
\end{figure}

Figure \ref{fig:m050z02} displays some evolutionary tracks of simple
$5 \msol$, $X=0.7,~Z=0.02$ model stars on the HR plane.  The continuous
black and red lines trace the evolution from the ZAMS to the base of the
first-giant branch making use of realistic microphysics as supplied by
{\tt MESA}. In particular, we made use of opacities from 
OPAL tables.  The full black line marks the part of the early evolution
when the QHE models are secularly stable. The red line, covers
the secularly unstable phase of evolution to the base of the giant branch. 
The full grey line traces the evolution of the same model star 
if $\Diff_t s \equiv 0$ is enforced, i.e. FE models. The latter
constraint gives numerical solutions only if the Henyey determinant
does not vanish, which happens if a star turns thermally unstable
via a monotonic secular eigenmode. At the epoch of the first
zero of the Henyey determinant, convergence fails as expected;
this point is highlighted as \emph{turning point of the sequence }on the
HR plane of Fig.~\ref{fig:m050z02}. 

The dashed lines illustrate the effect of evolving $5 \msol$,
$X=0.7,~Z=0.02$ model stars with
$\kappa(m) = \kappa_{\mathrm{e}^-}(m)$ only; the dashed black line
traces the respective QHE evolution. The grey dashed line on the other
hand traces the evolution of the FE case. QHE evolution to low effective 
temperatures stops and reverses at the onset of helium core-burning. The FE
evolution, in contrast, continues to very low temperatures, failing to
ignite helium core-burning. The lack of $H^-$ opacity in the $\kappa$
prescription lets the model star miss the Hayashi line so that it
continues to cool, always in full FE, until the code fails to converge
due to numerical problems in the equation of state. Most
importantly though, none of the computed FE $\mathrm{e}^-$-scattering
models did encounter a zero in their Henyey determinant, which means
that secular instability was indeed suppressed. This is compatible
with Murai's conjecture that the spatial $\kappa$ structure due to
Kramers opacity contribution is responsible for the onset of multiple
stellar models in equilibrium, or of secularly unstable QHE models,
which also applies to the Sch\"onberg-Chandrasekhar instability.

\begin{figure} 
	\begin{center}
	\includegraphics[width=0.90\textwidth]{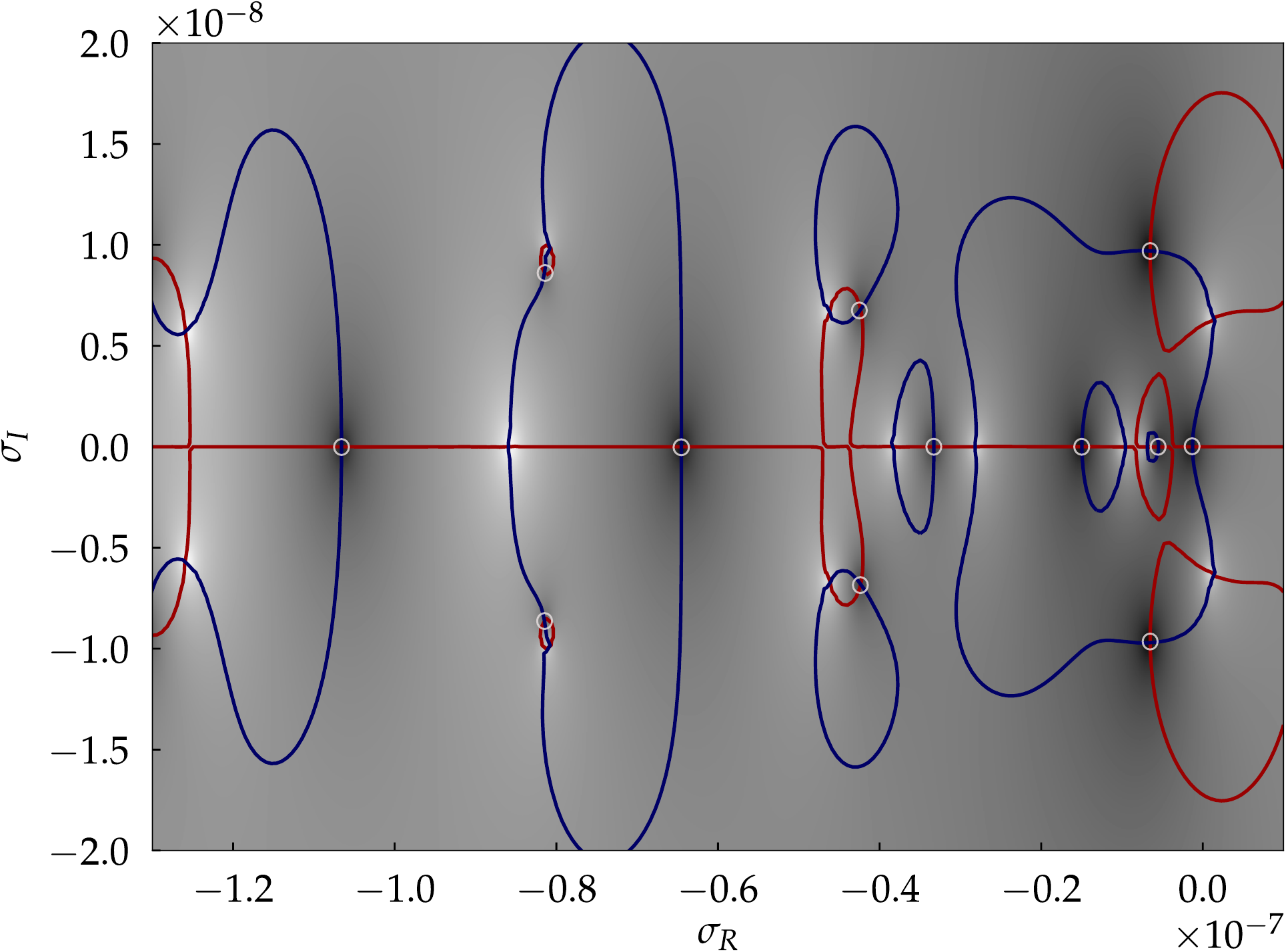} 
	\end{center}
     \caption{ Logarithm of the squared determinant, $D$, of the secular
     		   eigenproblem in the complex $\sigma$ plane of the $5 \msol$
     		   model of the above sequence, located at $\log\teff=4.15, 
     		   \log \llsol = 3.04$.
     		   Blue loci show Re$(D) = 0$, red loci trace Im$(D) = 0$. 
     		   The discrete dark spots~--~emphasized by white circles~--~on 
     		   the grey-scale map indicate the positions local minima 
     		   where secular \emph{eigenfrequencies }lie. Evidently, a mix of
     		   oscillatory and monotonous eigensolutions prevail.  
             } \label{fig:SecularCmplxPlane}
\end{figure}

To determine the range of secularly stable QHE models (the full black
line in Fig.~\ref{fig:m050z02}) we computed secular eigensolutions of
the respective stellar evolution models. For a $5 \msol$ QHE model at the
gate to the Hertzsprung-gap, at $\log\teff=4.15, \log\llsol=3.04$, a
pertinent region of the complex frequency, $\sigma = (\sigr,\sigi)$,
plane\sidenote{A time dependence $\propto \exp \sigma t$ was adopted.}
was scanned to visualize the distribution of the lowest-order secular
eigenmodes thereon.  Figure~\ref{fig:SecularCmplxPlane}
shows the frequency plane with $\sigma$ measured in units
of the stars' free-fall frequency.  The magnitude of the
determinant of the secular eigenproblem \citep[computed with the
Riccati method, cf.][]{alfalt07} is plotted as a grey-scale map. The
larger the magnitude of the determinant the brighter the
color.\sidenote[][-1cm]{The white patches usually hint at poles of the 
                   determinant function. The zeroes of the determinant
                   (i.e. the eigenfrequencies) are the black centers 
                   of the dark patches highlighted with white circles.}
Superimposed on the colormap are the loci of the zeroes of the real
part (blue) and of the imaginary part (red) of the determinant function. 
Local minima of this determinant function, indicating eigenfrequencies,
are highlighted with small white circles. Evidently, oscillatory and
monotonous modes coexist in the chosen region of the frequency
plane.  The three oscillatory modes appear twice on the complex plane
shown in Fig.~\ref{fig:SecularCmplxPlane}
because the secular problem is symmetric about $\sigi=0$.  As
mentioned further up, secular eigenvalues can be complex and
frequently are so, constituting \emph{oscillatory }eigenmodes; these
have been known since \citet{Schwarzschild1968}. \citet{Aizenman1971}
showed then that oscillatory secular modes occur even in rather simple
model stars close to the main sequence.

\subsection{B.2. A $15\,\msol$ star: FE versus NNE evolution}
\label{sec:FEorNNE}
\begin{figure} 
	\begin{center}
	  \includegraphics[width=0.99\textwidth]{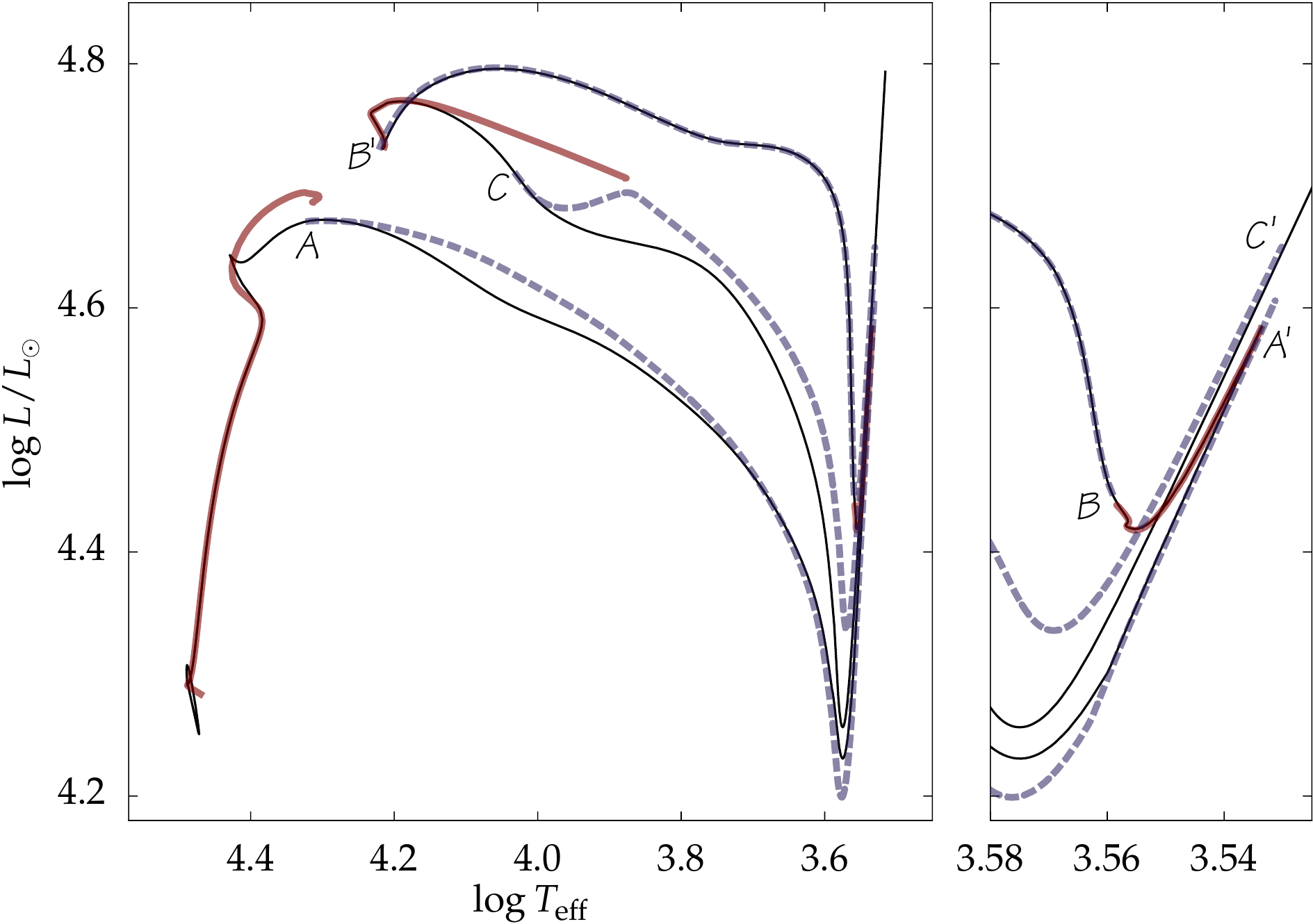}
	\end{center} 
	\caption{Evolutionary tracks of a $15 \msol$ star with
                 $X=0.7,Z=0.02$.  Superimposed on the thin black line of 
                 QHE evolution are phases of FE (red) and NNE
                 (indigo). The right panel is a zoom-in 
                 to more easily identify the different phases
                 in the red-giant region.}
    \label{fig:FEandNNE_m150}
\end{figure}

In the FE approximation, $\Diff_t s \equiv 0$ is inflicted upon a 
star's evolution. In the other extreme, the \emph{no-nuclear-evolution case }(NNE),  
$\Diff_t s$ is fully accounted for but any \emph{nuclear evolution }is 
neglected.\sidenote{In \texttt{MESA}, this can be achieved by setting
\texttt{dxdt{\_}nuc{\_}factor=0d0} in the \texttt{controls} namelist.}
Nuclear energy generation works as usual but it has 
no compositional consequences for the star in the NNE approximation. 
From Fig.~\ref{fig:AllBlueLoops} we learn that the $15 \msol$ model 
displayed there embarks on an extensive blue loop during helium core- and
double shell-burning.  Therefore, this $15 \msol$ model sequence is 
used to illustrate how conservative QHE evolution can be roughly pieced 
together as a succession of FE and NNE phases, at least during 
the H- and He-burning stages. Figure~\ref{fig:FEandNNE_m150} replots the 
$15 \msol$ QHE track of Fig.~\ref{fig:AllBlueLoops} as the continuous black line. 

The first FE episode extends from the ZAMS to the end of the S-bend 
(close to epoch $A$). The second FE phase starts at the tip of the first-giant
branch and it extends to the local luminosity minimum when the model stars 
embark on their blue loops (from epoch  $A'$ to $B$). The third FE stretch 
starts at the blue end of the loop (at epoch $B'$) where the model star starts 
to evolve back towards the red-giant region. FE evolution terminates 
at about $\log\teff=3.85$ when the model goes secularly unstable again. 
Compared with the QHE track, the first and the third FE phase are partially 
overluminous, this happens whenever QHE evolution experiences 
significant $\Diff_t s$ contributions, which is ignored in the FE case. 

The NNE computations could never be started using the last converging FE model. 
It always took a QHE model with sufficient $\Diff_t s$ contributions in the
neighborhood of a terminal FE model for an NNE sequence to launch successfully.
Therefore, the starting epochs and also their respective positions on the HR plane 
of the starts of NNE sequences are afflicted with some uncertainty. 
All the NNE phases, $A - A'$, $B - B'$, and $C - C'$, are lived through in 
a few $10^4$~years; i.e. on thermal timescales of the model stars.  
Notice that initial and final NNE epochs $A, A'$, $B, B'$, and $C, C'$, 
respectively, are pairs of models with the same chemical
compositions but different internal structures; they are hence 
examples of models violating the global VR theorem; a point made already by
\citet{Kaehler1978}. In other words, double solutions are regularly
encountered in canonical stellar-evolution computations under 
unspectacular circumstances. Once the final new equilibrium 
stratifications were established, the time steps adopted by the NNE computations 
grew rapidly because the structure of the model stars
did not change anymore. Much in contrast to what was reported at the beginning of 
this paragraph for terminal FE models, the final models of the NNE sequences 
served as reliable starting points for ensuing FE evolution 
(cf. epochs $A'$, $B'$, $C'$); usually lying nearby the 
respective QHE model on the HR plane. 

\bigskip

\newthought{Acknowledgments}: This work relied heavily on the
services of NASA's Astrophysics Data System. The do-them-yourself star models 
referred to in this exposition were computed with the {\tt MESA }program 
suite \citep[see e.g.][and references therein to earlier instrument
papers]{Paxton2015}. The stellar-evolution computations for 
this essay are so canonical that their reproduction does not
require a particular version of the {\tt MESA} package. I am indebted
to P.~Secchi and D.~Ducomet for advice and guidance to the
mathematical fluid-dynamics literature.

\bibliographystyle{aa} 
\bibliography{StarBase}

\end{document}